\documentclass[11pt,preprint]{aastex}
\usepackage{emulateapj5,psfig} 
\submitted{Received 2012 -- --; Accepted: 2012 -- --}

\def\gs{\mathrel{\raise0.35ex\hbox{$\scriptstyle >$}\kern-0.6em
\lower0.40ex\hbox{{$\scriptstyle \sim$}}}}
\def\ls{\mathrel{\raise0.35ex\hbox{$\scriptstyle <$}\kern-0.6em
\lower0.40ex\hbox{{$\scriptstyle \sim$}}}}

\newenvironment{inlinefigure}{%
\def\@captype{figure}%
\noindent\begin{minipage}{0.999\linewidth}\small}
{\end{minipage}\smallskip}
\makeatother

\lefthead{Smail et al.}
\righthead{Far-infrared powered Inverse Compton X-ray halos around Radio Galaxies}

\begin{document}

\title{Inverse Compton X-ray halos around high-z radio
  galaxies:\\ A feedback mechanism powered by far-infrared starbursts or the CMB?}

\author{
Ian Smail,$\!$\altaffilmark{1}
Katherine M.\ Blundell,$\!$\altaffilmark{2}
B.\,D.\ Lehmer\altaffilmark{3} \&
D.\,M.\ Alexander\altaffilmark{4}
}

\altaffiltext{1}{Institute for Computational Cosmology, Durham University, South Road, Durham DH1 3LE, UK}
\altaffiltext{2}{University of Oxford, Astrophysics, Keble Road, Oxford OX1 3RH, UK}
\altaffiltext{3}{The Johns Hopkins University, Homewood Campus, Baltimore, MD 21218, USA; NASA Goddard Space Flight Centre, Code 662, Greenbelt, MD 20771, USA}
\altaffiltext{4}{Department of Physics, Durham University, South Road, Durham DH1 3LE, UK}

\setcounter{footnote}{4}

\begin{abstract}
We report the detection of extended X-ray emission around two powerful radio galaxies at $z$\,$\sim$\,3.6 (4C\,03.24 and 4C\,19.71) and  use these  to investigate the origin of extended, Inverse Compton (IC) powered X-ray halos at high redshifts.  The halos have X-ray luminosities of  $L_{\rm X}\sim$\,3\,$\times$\,10$^{44}$\,erg\,s$^{-1}$ and sizes of $\sim $\,60\,kpc.  Their morphologies are broadly similar to the $\sim$\,60-kpc long radio lobes around these galaxies suggesting they are formed from IC scattering by relativistic electrons in the radio lobes, of either Cosmic Microwave Background (CMB) photons or far-infrared photons from the dust-obscured starbursts in these galaxies.  These observations double the number of $z$\,$>$\,3 radio galaxies with X-ray detected IC halos. We compare the IC X-ray to radio luminosity ratios for the two new detections to the two previously  detected $z$\,$\sim$\,3.8 radio galaxies. Given the similar redshifts, we would expect comparable X-ray IC luminosities if   millimeter photons from the CMB are the dominant seed field for the IC emission (assuming all four galaxies have similar ages and  jet powers).  Instead we see that the two $z$\,$\sim$\,3.6 radio galaxies, which are $\sim $\,4\,$\times$ fainter in the far-infrared than those at $z$\,$\sim $\,3.8,  also have  $\sim $\,4\,$\times$ fainter X-ray IC emission. Including data for a further six $z\gs $\,2 radio sources with detected IC X-ray halos from the literature, we  suggest that in the more compact, majority of radio sources, those with lobe sizes $\ls $\,100--200\,kpc,  the bulk of the IC emission may be driven by scattering of locally produced far-infrared photons from luminous, dust-obscured starbursts  within these galaxies, rather than millimeter photons from the CMB.  The resulting X-ray emission appears sufficient to ionise the gas on $\sim$\,100--200-kpc scales around these systems and thus help form the extended, kinematically-quiescent Ly$\alpha$ emission line halos found around some of these systems.  The starburst and AGN activity in these galaxies are thus combining to produce an even more effective and wide-spread ``feedback'' process, acting on the long-term gas reservoir for the galaxy, than either individually could achieve.  If episodic radio activity and co-eval starbursts are common in massive, high-redshift galaxies, then this IC-feedback mechanism may play a role in affecting the star-formation histories of the most massive galaxies at the present day.
\end{abstract}

\keywords{cosmology: observations --- galaxies: individual (4C\,03.24, 4C\,19.71) --- galaxies: evolution --- galaxies: formation }

\section{Introduction}

Galaxy formation models are increasingly using feedback from active galactic nuclei (AGN) to tune the evolution of star formation in their host galaxies (e.g.\ Bower et al.\ 2006, 2008; Sijacki et al.\ 2007; de Young 2010).  This is an effective way to counter the catastrophic over-cooling of gas (and hence excess star formation) in massive halos, which would otherwise result in overly luminous local ellipticals.  Feedback from AGN is independent of the traditional mechanical or radiation-driven mechanisms powered by star formation, and the resulting coupling of the growth of super-massive black holes (SMBH) and star formation may explain the correlation of stellar and SMBH masses seen in local spheroids (e.g.\ di Matteo et al.\ 2005).  However, if AGN feedback mechanisms are going to work it is critical that they have their maximum efficiency at the epoch where the bulk of the stars in the most luminous elliptical galaxies are formed, $z$\,$\geq $\,3--5 (e.g.\ Smith et al.\ 2012).

AGN feedback can influence the gas reservoirs in galaxies through a variety of radiative or mechanical mechanisms, and we focus here on the observational evidence which points to the importance of one such mechanism: Inverse Compton (IC) heating in massive galaxies at $z$\,$\sim $\,2--4. This evidence comes from the detection of extended X-ray emission around  high-redshift powerful radio galaxies (HzRG) and radio-loud quasars  (e.g.\ Carilli et al.\ 2002; Scharf et al.\ 2003; Fabian et al.\ 2003, 2009; Overzier et al.\ 2005; Blundell et al.\ 2006; Erlund et al.\ 2006, 2008; Smail et al.\ 2009; Laskar et al.\ 2010; Blundell \& Fabian 2011). This X-ray emission has been linked to IC scattering of millimeter photons (from the Cosmic Microwave Background, CMB), or potentially far-infrared (IR) photons, by electrons in the halos of these galaxies (see Mocz et al.\ 2011a,b).  The deepest of these exposures (e.g.\ the 210-ks {\it Chandra} integration on a $z$\,$\sim $\,1.8 radio galaxy by Fabian et al.\ 2003) yield unambiguous evidence that this extended emission has a non-thermal, power-law spectrum with a typical photon index of $\Gamma_{\rm eff}\sim $\,1.5--2, as expected from IC emission, rather than thermal emission from hot gas trapped in a potential well.

Detections of the most distant extended X-ray emission come from the 130-ks and 100-ks {\it Chandra X-ray Observatory} observations of two far-infrared-luminous HzRGs:  4C\,41.17 and 4C\,60.07 (Scharf et al.\ 2003; Smail et al.\ 2009), both at $z$\,$=$\,3.8.   These X-ray observations show non-thermal emission  with luminosities of $\sim$\,10$^{45}$\,erg\,s$^{-1}$  extending over $\sim $\,100-kpc (12--15$''$) scales around both HzRGs. This emission  most likely arises from IC scattering of far-infrared/millimeter  photons by the electron population within the radio lobes (Scharf et al.\ 2003).  However, these photons come from very different sources: the millimeter photons come from the CMB, which is some $\sim $\,500\,$\times$ more intense at $z$\,$=$\,3.8 than $z$\,$=$\,0, whereas the far-infrared photons come from the starbursts in these HzRGs, which have total infrared luminosities of $L_{\rm IR}\sim $\,10$^{13}$\,L$_\odot$, comparable to their IC X-ray luminosities (Stevens et al.\ 2003; Ivison et al.\ 2008).   

The possibility that there are two sources of photons driving the IC X-ray emission around HzRGs may make this process more influential in the evolution of massive galaxies.  Indeed, in addition to their extended X-ray emission, both 4C\,41.17 and 4C\,60.07 also exhibit very luminous and extended Ly$\alpha$ halos (van Breugel et al.\ 1998), which are roughly co-aligned with their X-ray halos and radio lobes.\footnote{We make a distinction between the formation of the large-scale Ly$\alpha$ halos, which are typically kinematically quiescent (e.g.\ Villar-Martin et al.\ 2003), and the generally smaller-scale, higher surface brightness and typically kinematically more disturbed ionised structures which are formed by shock heating by the passage of the jets (e.g.\ Best et al.\ 2000).  }  These extended, kinematically-quiescent Ly$\alpha$ halos may be an indication of significant heating from the starburst or AGN within these systems and so are a signature of ``feedback'' processes which are thought to influence the evolution of massive galaxies and SMBHs at high redshifts (their luminosities also appear to evolve rapidly with redshift, Zirm et al.\ 2009). In fact, it appears that the IC-generated X-ray halos would be sufficient to photo-ionize the extended Ly$\alpha$ halos (Scharf et al.\ 2003)  making this a {\it new} and previously unappreciated feedback mechanism with three key features: 1) it affects the gas across the whole of the galaxy halo, $\sim$\,100\,kpc; 2) it operates preferentially in the most massive halos (those halos with SMBHs large enough to drive powerful radio lobes); 3) it is most active at the highest redshifts (owing to the rapid growth with redshift of both the CMB energy density and total infrared luminosities of HzRGs; Archibald et al.\ 2001).  The mass-specific nature and redshift-dependency of this feedback mechanism may thus provide a new avenue for theoretical attempts to model the growth of the most massive galaxies seen in the local Universe (Bower et al.\ 2006).

IC-powered X-ray halos may  be a common feature of HzRGs (Celotti \& Fabian 2004; Mocz et al.\ 2011a,b), however, hitherto only two HzRGs at $z$\,$>$\,3 have the deep {\it Chandra} observations necessary to identify IC emission: 4C\,41.17 and 4C\,60.07.  These galaxies have similar sized radio lobes, as well as {\it identical} redshifts (and hence CMB backgrounds), {\it identical} total infrared luminosities and they were found to have {\it identical} IC luminosities (Smail et al.\ 2009). To increase the number of high-redshift IC detections and to attempt to identify the dominant photon source for the IC emission we have therefore obtained sensitive {\it Chandra} X-ray observations of a further two HzRGs, 4C\,03.24 ($z=$\,3.57) and 4C\,19.71 ($z$\,$=$\,3.59). These are close to the redshifts of 4C\,41.17 and 4C\,60.07, and so have comparable CMB backgrounds and they also have similar radio luminosities and lobe sizes to the previously studied HzRGs, but much lower total infrared luminosities.  By comparing the IC X-ray emission  around  the four HzRGs we hope to distinguish the origin of the dominant IC seed photon field: millimeter photons from the CMB or far-infrared photons from a starburst. This target selection is critical to understand both the evolution of the IC emission process with redshift and the importance of this AGN feedback mechanism in the formation of massive galaxies.

In the next section we describe the {\it Chandra} X-ray observations of 4C\,03.24 and 4C\,19.71, and their reduction and analysis.  In \S3 we present and discuss our results and in \S4 we give our conclusions. In our analysis we assume a cosmology with $\Omega_{\rm M}$\,$=$\,0.27, $\Omega_\Lambda$\,$=$\,0.73 and $H_0$\,$=$\,71\,km\,s$^{-1}$\,Mpc$^{-1}$, giving an angular scale of 7.4\,kpc\,arcsec$^{-1}$ at $z\sim$\,3.6.

%
%
\begin{figure*}[tbh]
\centerline{\psfig{file=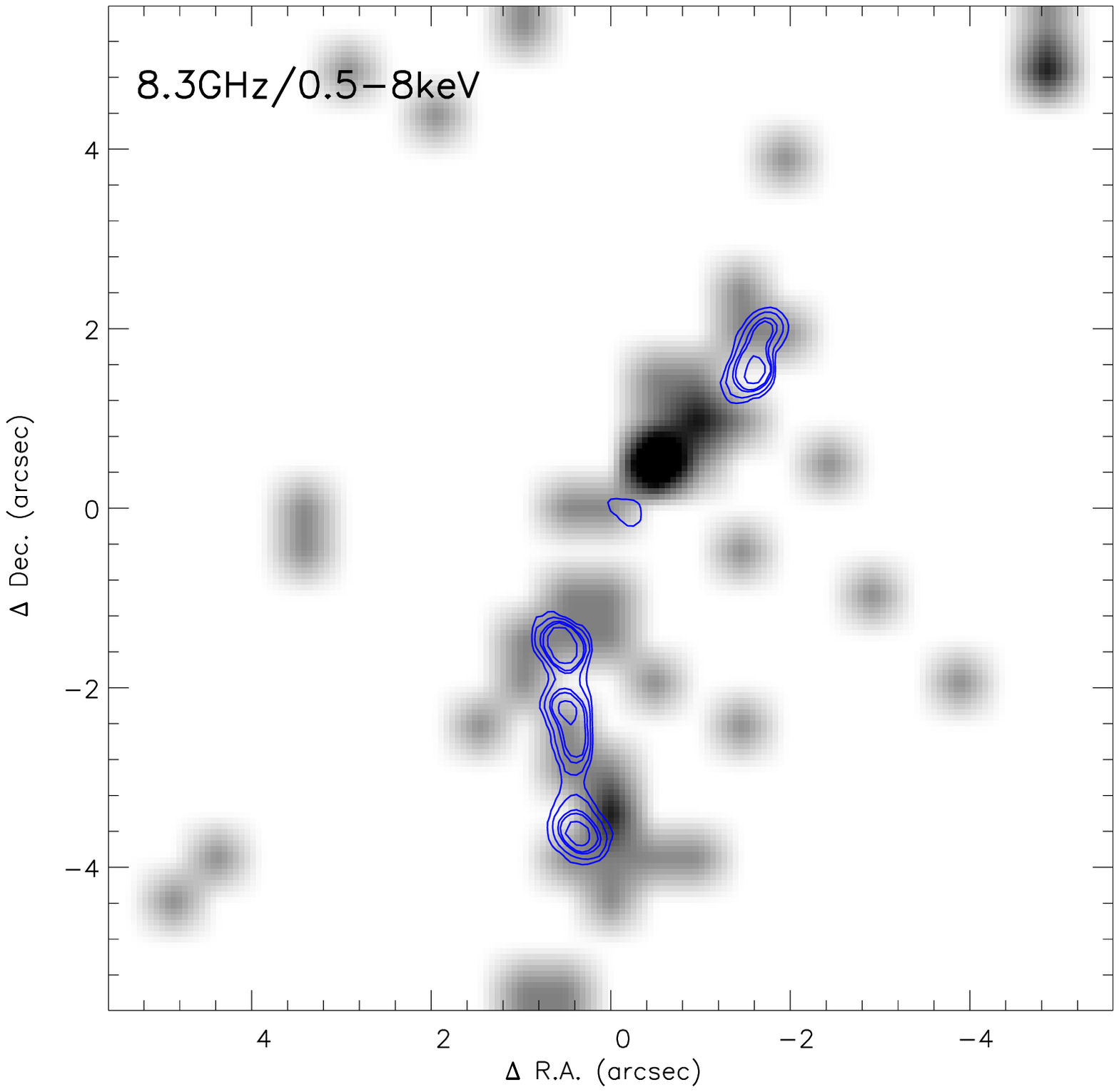,width=3.2in,angle=0} \psfig{file=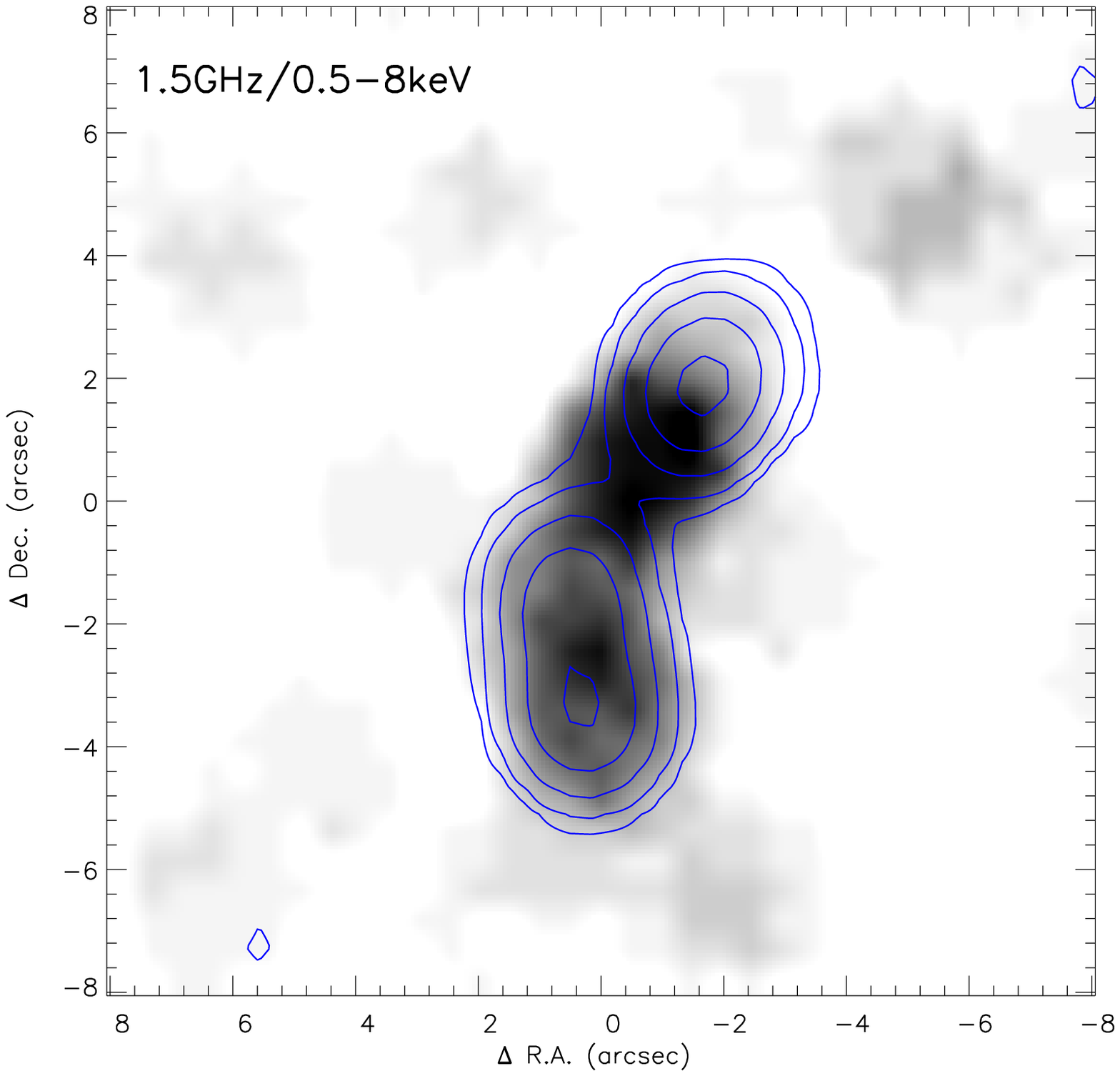,width=3.2in,angle=0}}
\vspace*{-0.5cm}
\centerline{\psfig{file=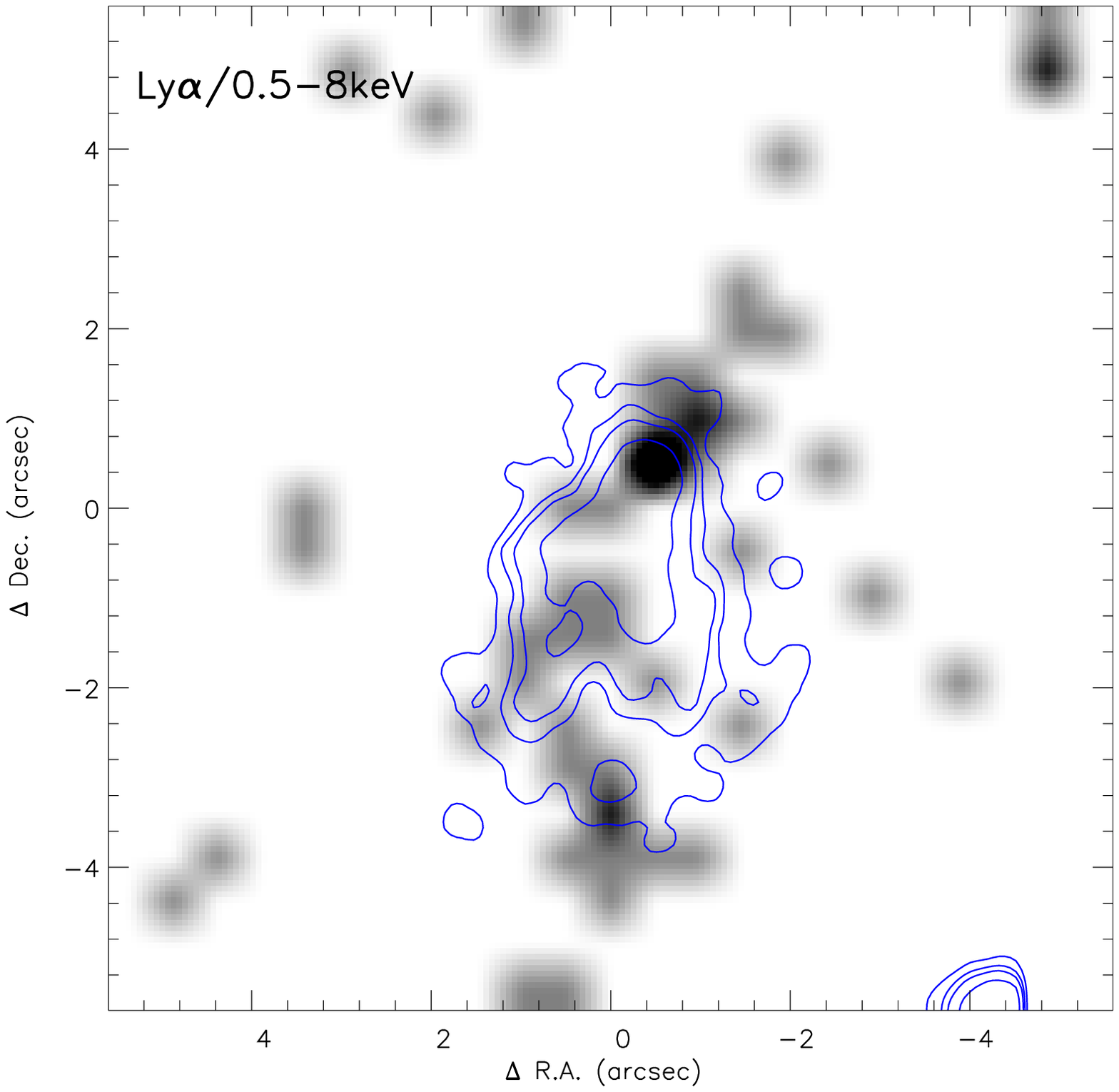,width=3.2in,angle=0}
\psfig{file=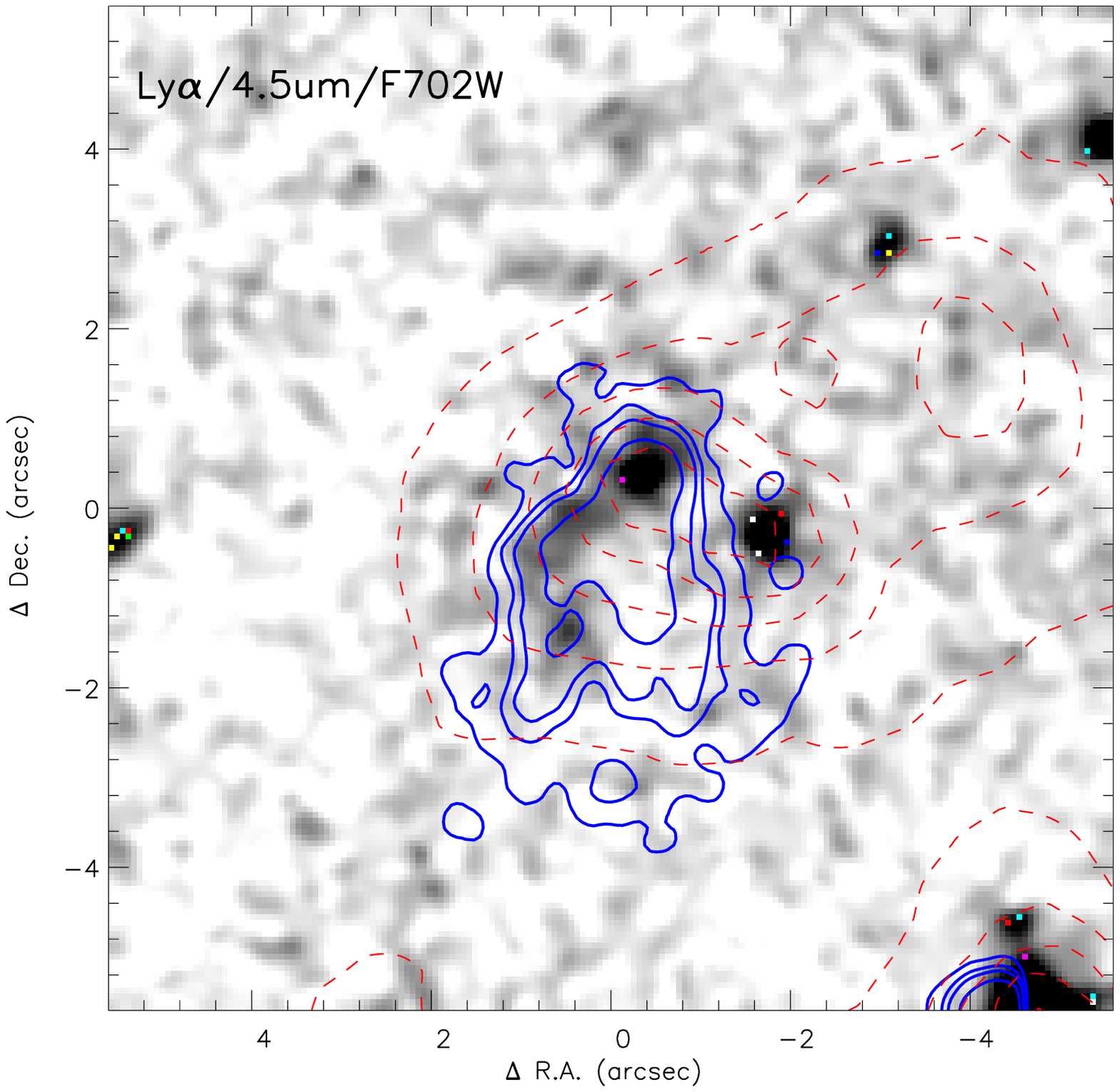,width=3.2in,angle=0}}
\vspace*{-0.3cm}
\caption{\small Four views of the field around 4C\,03.24: 
{\it (upper-left)} the VLA 8.3-GHz map contoured over the 0.5--8\,KeV {\it Chandra} image, the latter is smoothed with a 0.5-arcsec Gaussian kernel for display purposes;
{\it (upper-right)} a slightly expanded view of the VLA 1.5-GHz map contoured over the smoothed 0.5--8\,KeV {\it Chandra} image (using a 1.5$''\times$\,1.5$''$ top hat smoothing kernel to match the radio and X-ray resolutions); 
{\it (lower-left)} the Ly$\alpha$ emission contoured over the 0.5--8\,KeV {\it Chandra} image; 
{\it (lower-right)}  A {\it HST} WFPC2 F702W image of the HzRG (grayscale) with the Ly$\alpha$ emission (solid, van Ojik et al.\ 1996) and 4.5-$\mu$m {\it SST} IRAC map (dashed) overlayed as contours.  These images demonstrate that the very weak, but detectable, extended X-ray emission around 4C\,03.24 is extended in the same direction as the radio lobes in the galaxy.  We also see weak emission which may come from the core of the radio galaxy (visible as a point-source in the {\it HST} image). The Ly$\alpha$ emission is only detected to the south of the core, where the correspondance between the X-ray and radio emission is also closest, but has an extent which is comparable to the X-ray emission.  Finally, a comparison of the F702W and IRAC images shows that the radio galaxy has two very nearby companions, one with similar $R_{702}-4.5\mu$m colors to 4C\,03.24 and the other which is very red.}
 \end{figure*}

%
%
\begin{figure*}[tbh]
\centerline{\psfig{file=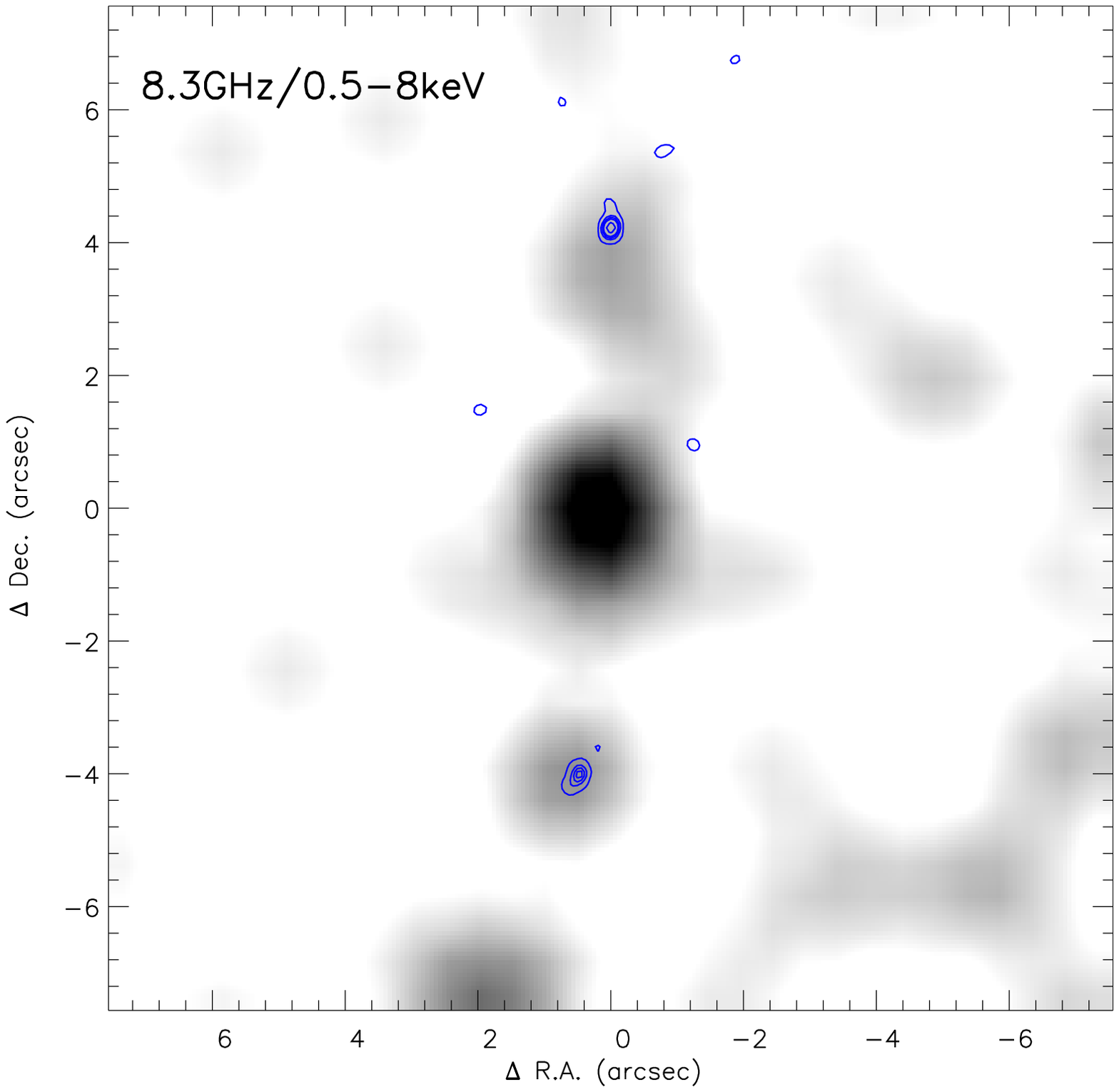,width=3.2in,angle=0} \psfig{file=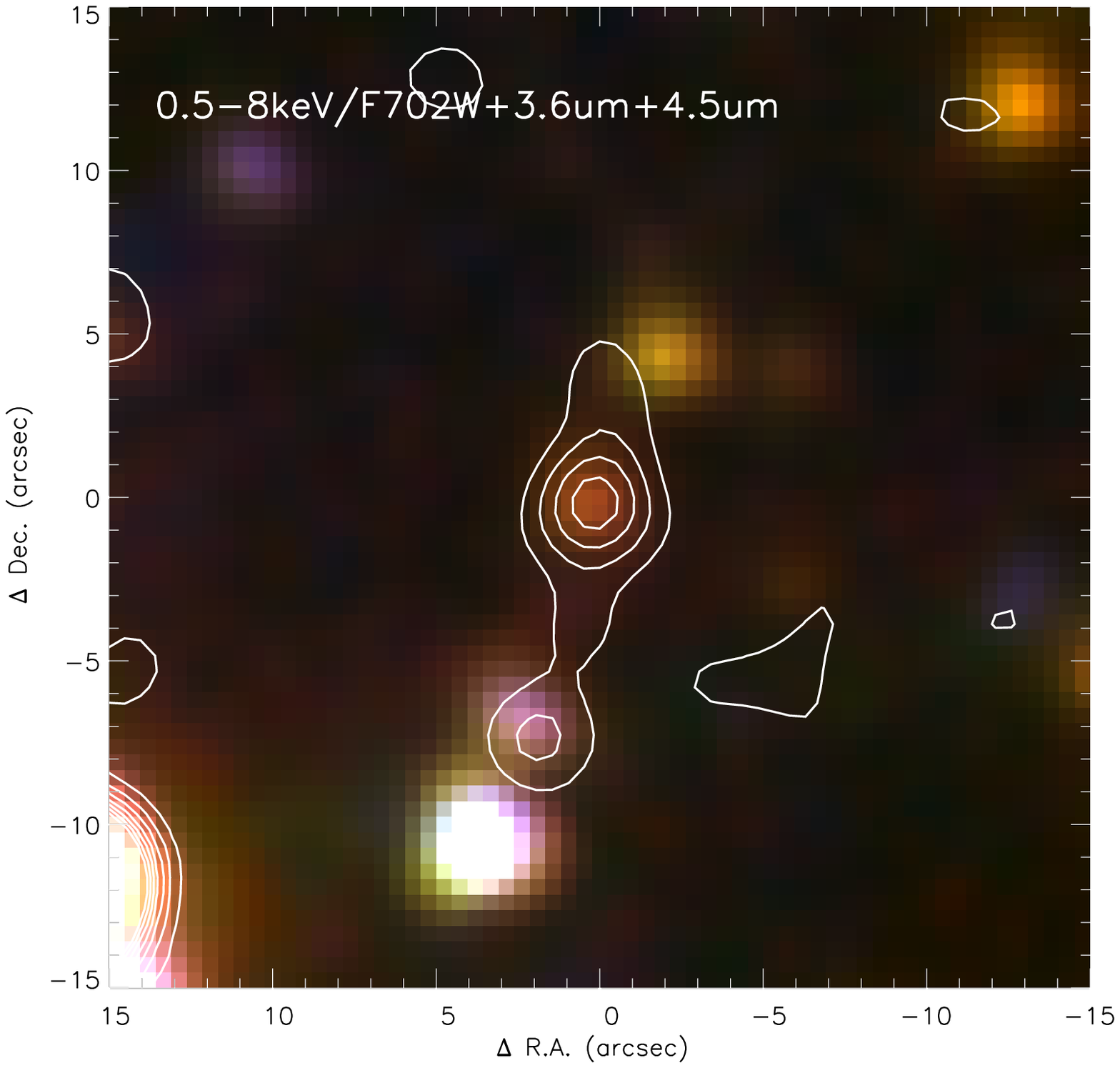,width=3.2in,angle=0}}
\vspace*{-0.5cm}
\centerline{\psfig{file=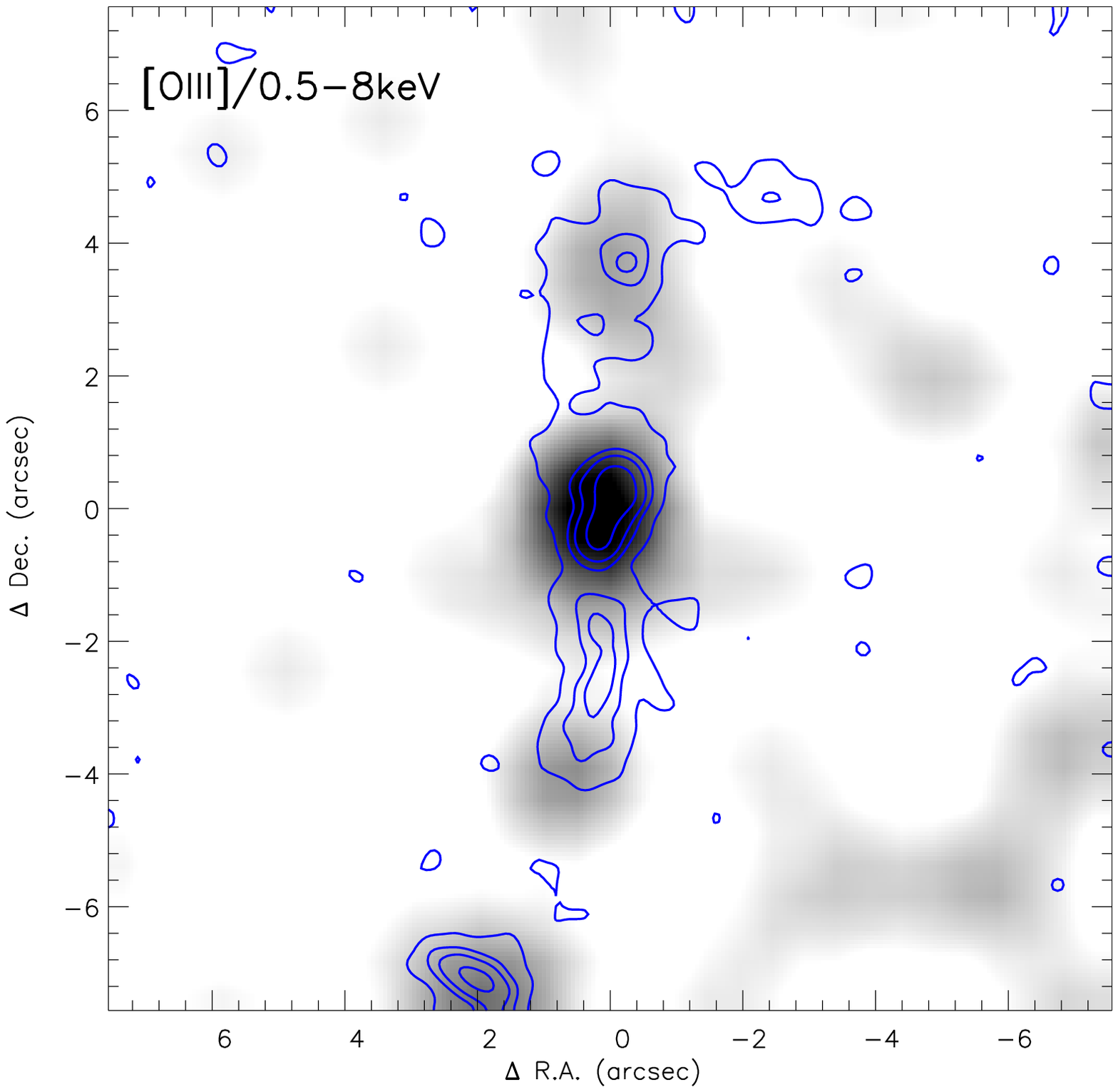,width=3.2in,angle=0} \psfig{file=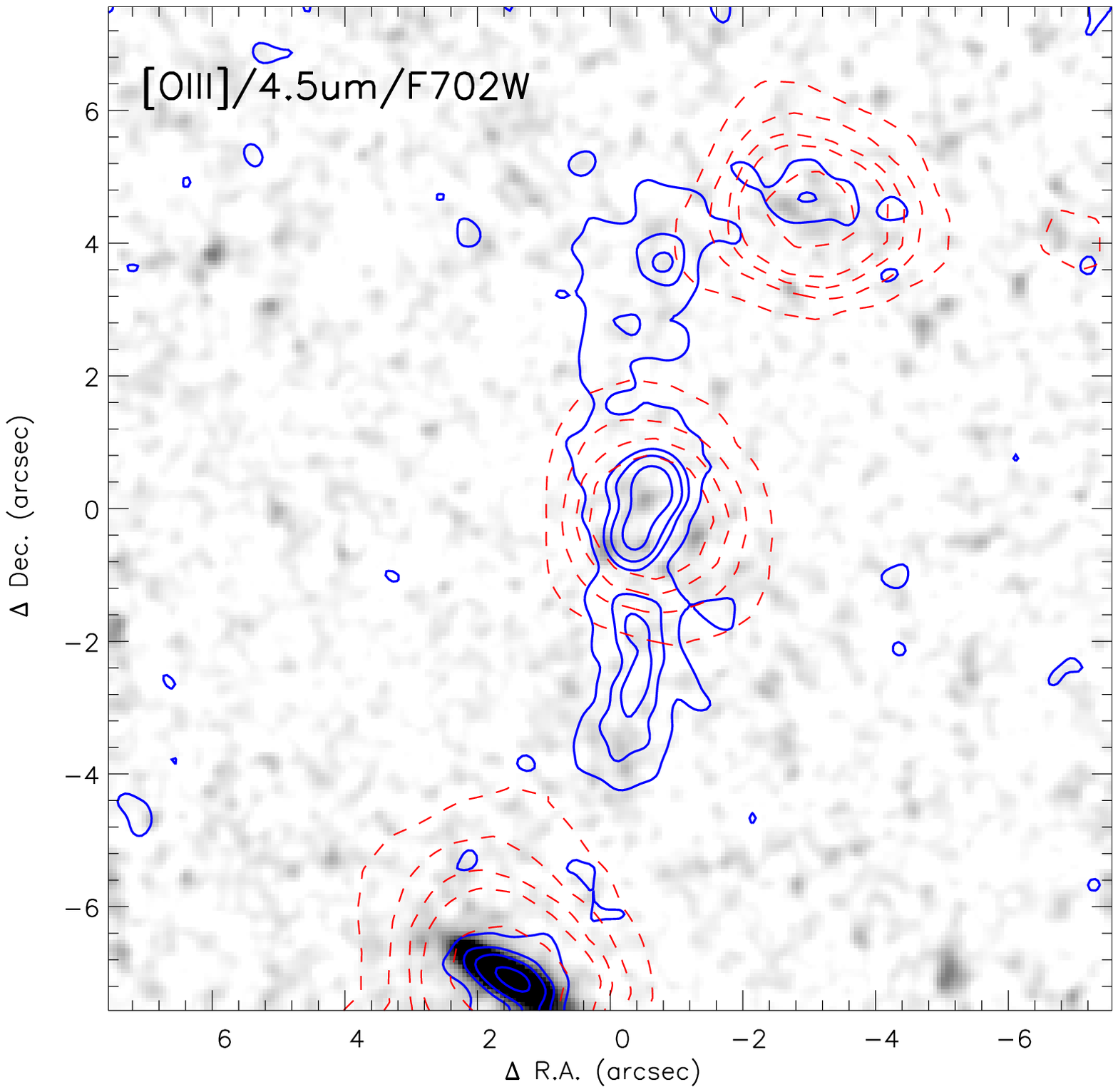,width=3.2in,angle=0}}
\vspace*{-0.3cm}
\caption{\small Four views of the field around 4C\,19.71: 
{\it (upper-left)} the VLA 8.3-GHz map contoured over the smoothed 0.5--8\,KeV {\it Chandra} image (using a 1.5$''\times$\,1.5$''$ top hat smoothing kernel, to enhance the visibility of the very faint emission); 
{\it (upper-right)} an expanded view of the smoothed 0.5--8\,KeV {\it Chandra} image contoured over a true-color image constructed from {\it HST} WFPC2 F702W and  3.6-$\mu$m and 4.5-$\mu$m {\it SST} IRAC maps;  
{\it (lower-left)} the [O{\sc iii}]\,5007 emission (Armus et al.\ 1998) contoured over the smoothed 0.5--8\,KeV {\it Chandra} image, showing the presence of an ionised gas halo around the HzRG (unfortunately no Ly$\alpha$ map is available for this target); 
{\it (lower-right)}  An {\it HST} WFPC2 F702W image of the HzRG (grayscale) with the [O{\sc iii}]\,5007 emission (solid) and 4.5-$\mu$m {\it SST} IRAC map (dashed) overlayed as contours.   Note the [O{\sc iii}] image is not continuum-corrected.  As with 4C\,03.24, these images demonstrate that the very weak, but detectable, X-ray emission is seen around the two lobes of the radio emission in 4C\,19.71.  The [O{\sc iii}] emission is also seen along the full extent of the lobes.  Finally, as with 4C\,03.24,  a comparison of the F702W and IRAC images shows that the radio galaxy has a red, close companion to the north. }
\end{figure*}

\section{Observations, Reduction and Analysis}

The two high-redshift radio galaxies studied here are: 4C\,03.24 (12\,45\,38.36, +03\,23\,20.7, J2000; MRC\,1243+036) at $z=$\,3.57 and 4C\,19.71 (21\,44\,07.48, +19\,29\,15.4, J2000; MG\,2141+192) at $z=$\,3.59.  These targets  were selected as they have similar radio luminosities, lobe sizes and redshifts to the previously studied 4C\,41.17 (Scharf et al.\ 2003) and 4C\,60.07 (Smail et al.\ 2009), see Tables 1 \& 2. Also, as with the two previously studied $z=$\,3.8 HzRGs, both galaxies show extended ionised halos in Ly$\alpha$ (van Ojik et al.\ 1996; Maxfield et al.\ 2002; Table~1) and in the case of 4C\,19.71 also [O{\sc iii}]\,5007 emission (Armus et al.\ 1998).  These two galaxies differ from the two previously-studied examples in terms of their sub-millimeter fluxes and hence their total infrared luminosities $L_{\rm IR}$ (8--1000\,$\mu$m): the measured 850-$\mu$m fluxes are S$_{\rm 850\mu m}=$\,2.3 and 4.6\,mJy for 4C\,03.24 and 4C\,19.71 respectively, much fainter than the 17 and 12\,mJy measured for 4C\,60.07 and 4C\,41.17 (Table~2; Archibald et al.\ 2001).  The corresponding total infrared luminosities are 3.2\,$\times$\,10$^{12}$ and 6.4\,$\times$\,10$^{12}$\,L$_\odot$ for 4C\,03.24 and 4C\,19.71, compared to 10\,$\times$\,10$^{12}$ and 16\,$\times$\,10$^{12}$\,L$_\odot$ for 4C\,60.07 and 4C\,41.17, respectively (Table~1).  These $L_{\rm IR}$ values assume a fixed dust temperature of $T_{\rm d}=$\,45\,K and $\beta=$\,1.5, as derived from a fit to the extensive far-infrared/millimetre photometry for 4C\,41.17 in Greve et al.\ (2007).  The faint total infrared luminosities of 4C\,03.24 and 4C\,19.71 are confirmed by their non-detections at 250, 350 and 500$\mu$m in observations with the SPIRE far-infrared camera on the {\it Herschel  Space Observatory} (N.\ Seymour, priv.\ comm.).  We also list in Table~1 the observed frame 1.7-GHz radio luminosities for all four sources from the fluxes listed by NED.  The similar redshifts, radio luminosities and extents, but fainter total infrared luminosities of 4C\,03.24 and 4C\,19.71 compared to 4C\,60.07 and 4C\,41.17, mean they are an ideal test of the relative contributions of local far-infrared and CMB photons as the seed population for IC emission around HzRGs.

The X-ray observations of these two galaxies were obtained with {\it  Chandra} whose high angular resolution and high sensitivity are essential to resolve any faint, extended X-ray emission around the HzRGs. We obtained deep {\it Chandra} ACIS-I pointings of 16.9$' \times $\,16.9$'$ regions centered close to 4C\,03.24 and 4C\,19.71. A 92.0-ks observation of 4C\,03.24 (Obs-IDs 12288) was obtained on 2010 December 05--06 and a total of 91.7-ks of observations of 4C\,19.71 between 2010 August 23--27 (36.5\,ks, Obs-ID 12287; 55.2\,ks, Obs-ID 13024).

The {\it Chandra} X-ray Center pipeline software (version 7.6.11) was used for the basic data processing.  As with Smail et al.\ (2009), the reduction and analysis followed Luo et al.\ (2008), including replacing the standard bad-pixel file with one which includes the obvious bad columns and pixels above 1\,KeV, which then excludes just $\sim $\,1.5\% of the total effective area. We inspected light-curves for all three observations and found no significant (i.e.\ $> $\,3$\sigma$ above the mean count rate) flaring events.  We then registered the observations by first running {\sc wavdetect} to generate an initial source list and matching these to counterparts in the 4.5-$\mu$m {\it  Spitzer Space Telescope} ({\it SST}\,) IRAC images of each field which are aligned to FK5 (these data come from the {\it SST} archive: 4C\,03.24, PID 50032; 4C\,19.71, PID 3329). This confirmed the absolute astrometry of both images is good to $\sim $\,0.4$''$. Finally we constructed images and exposure maps for both fields using the standard {\it ASCA} grade set ({\it ASCA} grades 0, 2, 3, 4, 6) for three standard bands: 0.5--8.0\,KeV (full band), 0.5--2.0\,KeV (soft band), and 2--8\,KeV (hard band).  The final, vignetting-corrected on-source effective exposure times at the positions of the radio galaxies are 88.1\,ks and 87.9\,ks for 4C\,03.24 and 4C\,19.71, respectively.

We show the resulting X-ray images for 4C\,03.24 and 4C\,19.71 in Figs.~1 and 2, respectively, and  compare the X-ray emission to the radio and optical/mid-infrared data. The 1.5-GHz radio map of 4C\,03.24 and 8.3-GHz map of 4C\,19.71 shown here were retrieved from the  VLA archive, while the 8.3-GHz map of 4C\,03.24 is from van Ojik et al.\ (1996). The optical data comes from the {\it Hubble Space Telescope} ({\it HST}\,) F702W WFPC2 imaging (PID: 6632) and has been smoothed with a 0.5$''$ FWHM Gaussian kernel to match the resolution of the {\it Chandra} data and increase the visibility of faint features.  We also show the morphology of the ionised halos around the two galaxies: Ly$\alpha$ for 4C\,03.24 from van Ojik et al.\ (1996) and [O{\sc iii}]\,5007 for 4C\,19.71 from Armus et al.\ (1998).

As Figs.~1 and 2 illustrate, both HzRGs appear to have faint extended X-ray emission which is coincident with their radio emission.  For 4C\,03.24 the emission extends south $\sim$\,40\,kpc from the core, with a northern extension of $\sim$\,20\,kpc, mirroring the asymmetry in the radio emission.  We measure 39$^{+10}_{-9}$ net counts in the 0.5--8-KeV band within a conservative 20$''$-diameter (150-kpc) aperture around 4C\,03.24, corresponding to a  count rate of 4.5$^{+1.1}_{-1.1}\times $\,10$^{-4}$\,cts\,s$^{-1}$.  The emission in 4C\,19.71 is weaker, with faint X-ray emission coincident with the radio hotspots at a radius of $\sim$\,30\,kpc, yielding 15$^{+6}_{-5}$ net counts in a 10$''$-diameter (75\,kpc) aperture and a corresponding count rate of 1.7$^{+0.7}_{-0.5}\times 10^{-4}$\,cts\,s$^{-1}$. In both cases the background count rates and their uncertainties were derived from randomly-placed apertures across each image and we also mask emission from the cores of the HzRGs by interpolation.\footnote{This core emission is only slightly brighter than the extended emission, with 9 net counts in 4C\,03.24 and 6 net counts in 4C\,19.71, so it is possible that it represents IC-powered X-ray emission from the inner regions of these galaxies.  If we included this emission, then the IC X-ray luminosities would increase by approximately the  1-$\sigma$ error.}  These count rates yield observed 0.5--8-KeV fluxes of 2.9$^{+0.8}_{-0.7}$\,$\times$\,10$^{-15}$\,erg\,s$^{-1}$ and 3.2$^{+1.3}_{-1.0}$\,$\times$\,10$^{-15}$\,erg\,s$^{-1}$, with 2--8\,KeV/0.5--2\,KeV band ratios of 0.4$\pm 0.3$ and 1.0$^{+0.9}_{-0.6}$ for 4C\,03.24 and 4C\,19.71 respectively.  The corresponding luminosities and effective photon indices are listed in Table~1.  This table also provides a summary of the properties of the two other $z>$\,3 HzRGs with detected IC halos: 4C\,41.17 and 4C\,60.07 from Scharf et al.\ (2003) and Smail et al.\ (2009) respectively.

\section{Results and Discussion}

Our observations confirm the presence of weak X-ray emission on $\sim$\,60\,kpc scales around 4C\,03.24 and 4C\,19.71, with morphological similarities to both their radio lobes and extended emission-line halos.  This spatial correlation between the extended X-ray and radio fluxes suggests that, as in 4C\,41.17 and 4C\,60.07, the X-ray emission most likely arises from IC scattering of electrons in the radio plasma off millimeter or far-infrared photons, associated  with either the CMB or the obscured starburst in these galaxies.  

These new observations  double the number of IC X-ray halos detected around HzRGs at $z$\,$>$\,3.  However, as Table~1 shows, the X-ray emission around both 4C\,03.24 and 4C\,19.71 is  $\sim $\,4\,$\times$ fainter than around the two previously detected HzRGs: 4C\,41.17 and 4C\,60.07. The two new HzRGs were chosen to closely match the redshifts (hence have comparable CMB energy densities), radio luminosities and lobe extents (so potentially comparable radio source ages) of 4C\,41.17 and 4C\,60.07, but to also have  $\sim $\,4\,$\times$ fainter total infrared luminosities (Table~1).   The difference between the apparent X-ray luminosities of the two sets of HzRGs suggests a possible role for far-infrared photons in powering the IC in the more luminous X-ray halos around 4C\,41.17 and 4C\,60.07.  Indeed, as we discuss below, the association of X-ray emission with just the hot-spots in 4C\,19.71, may be indicating that the bulk of the IC emission in this system is driven by scattering of CMB photons (an alternative explanation for the hotspot X-ray emission is synchrotron emission from the inner hotspot shock, as seen in the $z$\,$\sim$\,0.1 radio quasar 4C\,74.26, Erlund et al.\ 2010).  By comparison, therefore, the bulk of the emission in the much more X-ray luminous HzRGs may relate to scattering of far-infrared photons.

To quantify our analysis and broaden its scope, we have compiled a sample of other high-redshift radio galaxies or radio quasars with sensitive X-ray observations which reveal extended X-ray emission  which is likely to arise from IC.  We stress that this sample is in no way ``complete'', although it does represent all of the deep ($\gs $\,100\,ks) {\it Chandra} observations of $z\gs $\,1.5 radio galaxies of which we are aware.  We list these literature sources in Table~2 along with the observed 0.5--8\,KeV, 408-MHz and 850-$\mu$m fluxes and radio lobe extent (which is similar in size to the IC X-ray halos in all cases).  We convert the X-ray and radio fluxes to restframe 0.5--8\,KeV and 408-MHz luminosities using our adopted cosmology and the relationships from Alexander et al.\ (2003) with a photon-index of $\Gamma_{\rm eff}=$\,2 in the X-ray and a power-law index of $\alpha=-$1.2 in the radio (see below).  To derive total infrared luminosities (8--1000\,$\mu$m) we take the measured 850-$\mu$m sub-millimeter fluxes and assume a modified black body spectral energy distribution with $\beta=$\,1.5 and a characteristic dust temperature of T$_{\rm d}=$\,45\,K (see \S2).  In our analysis we will also differentiate the radio sources on the basis of their lobe, or equivalently IC X-ray halo, sizes into small ($<$\,100\,kpc), intermediate (100--500\,kpc) and large ($\gs$\,500\,kpc) using  the lobe sizes listed in Table~2.  This is motivated by the expectation that the contribution of far-infrared driven IC emission in the largest radio sources, $\gs$\,500\,kpc, is likely to be negligible (e.g.\ Laskar et al.\ 2010), and so we will seek differences in the IC properties of these systems, compared to the more compact radio sources.

We wish to test whether either, or both, of the CMB and the obscured starbursts in these high-redshift radio sources could provide the source of sub-millimeter/far-infrared photons to scatter off the relativistic electron population in their radio lobes.  The relationship for IC scattering connects the energy of the resulting IC X-ray photons to the frequency of the input far-infrared or sub-millimeter photons and the energy of the relativistic electrons via: 

$$E \sim \,0.09\, T\nu / B \sim 3.8\times 10^{-7} T \gamma^2$$ 

where $E$ is in KeV, $T$ is the characteristic temperature of the photons in K, $B$ is the magnetic field strength in $\mu$G, $\nu$ is the synchrotron frequency of the electrons in MHz and $\gamma$ is their corresponding Lorentz factor (Felten \& Morrison 1966). 

We require electrons with $\gamma\sim $\,1500 in a radio lobe at $z\sim$\,3.6--3.8 to scatter CMB photons up to $\sim$\,9\,KeV (or 2\,KeV in our observed frame). These electrons will also emit synchrotron radiation at $\sim $\,400\,MHz (an observed frequency of $\sim $\,100\,MHz), assuming a magnetic field of $\sim $\,50$\mu$G  (see Scharf et al.\ 2003). At the same time, the more numerous electrons with $\gamma\sim$\,1000 will scatter far-infrared photons from a 45\,K black body to an observed-frame energy of 2\,KeV, and similarly radiate synchrotron emission at an observed frequency of $\sim$\,30\,MHz.     If only synchrotron losses are considered, then the radiative lifetime of these particles would be 5\,Myrs and 10\,Myrs for $\gamma\sim$\,1500 and $\gamma\sim$\,1000 electrons and our choice of $B$.  In reality, of course, the losses due to adiabatic expansion of the synchrotron plasma are particularly dramatic, as is the energy depletion of the synchrotron electrons from the IC process (Blundell \& Rawlings 2000). If instead we adopted a magnetic field of $\sim $\,10$\mu$G, the synchrotron emission from these   electrons would peak at   restframe frequencies of $\sim$\,100\,MHz and $\sim$\,30\,MHz respectively, and their synchrotron-cooling lifetimes would increase by $\sim$\,25\,$\times$. Thus the  densities (and spatial distributions) of the CMB and far-infrared photon fields, combined with the relative numbers and distribution of electrons in the lobes with $\gamma\sim $\,1000s (which are traced by the $\sim$\,0.1\,MHz radio emission) will determine the relative contributions of CMB and far-infrared scattering to the creation of (observed-frame) KeV-energy X-ray photons.

%
%
 \begin{inlinefigure}\vspace{6pt}
 \centerline{\psfig{file=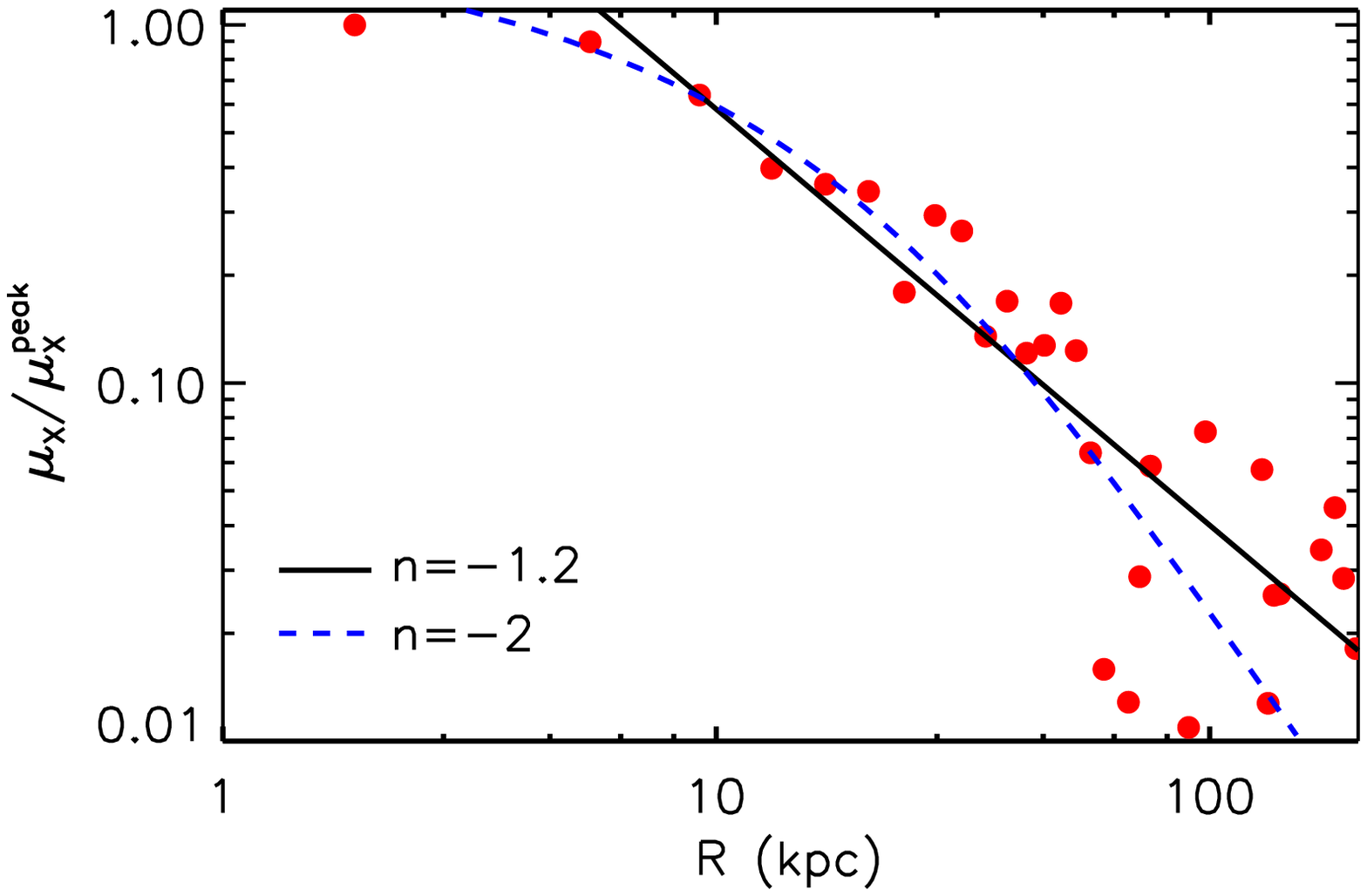,width=3.5in,angle=0}}
 \vspace{6pt}
 \noindent{\small {\small \sc Fig.~3.--- }
The normalised, average X-ray surface brightness profile derived by stacking the extended X-ray emission around 4C\,03.24, 4C\,19.71, 4C\,60.07 and 4C\,41.17.  We overplot the best-fit power law  with $\mu_X\propto R^n$ with $n\sim -$1.2, as well as the expected profile from a toy model of a 10:3 axial ratio  lobe illuminated by an  $R^{-2}$ photon field.  We conclude that the average X-ray surface brightness profile of the IC halos around these HzRGs is consistent with toy models comprising either narrow lobes illuminated by a photon field  which is more extended than a central point source, or relatively ``fat'' lobes illuminated by a central source. Hence the depth of the current X-ray observations of these $z\sim$\,3.6--3.8 HzRGs is insufficient to reliably distinguish the spatial form of the photon field responsible for their IC emission. } 
 \end{inlinefigure}

The first quantitative test of the IC hypothesis for our sample is to look at the photon index of the X-ray emission. The IC photon index is related to the power-law index of the radio emission for electrons having the same Lorentz factors, $\alpha$, by $\Gamma_{\rm eff}=1-\alpha$.  We estimate $\alpha\sim -$1.2\,$\pm$\,0.2 for the four $z>$\,3 HzRGs at observed frequencies between 400\,MHz and 1.4\,GHz, which roughly match the frequency range of the synchrotron emission from electrons with the necessary Lorentz factors to produce the IC X-ray emission. Using this mean radio spectral index  we predict an X-ray photon index: $\Gamma_{\rm  eff}=$\,2.2\,$\pm$\,0.2.  By comparison, the effective photon indices of the X-ray emission around 4C\,03.24 and 4C\,19.71 are 1.8$^{+0.7}_{-0.5}$ and 1.1$^{+0.9}_{-0.6}$ respectively, although neither is well-constrained due to the low significance of the detections and so they are formally consistent with the predicted value (Table~1).   Taking the photon indices for all ten high-redshift radio sources in our compilation, we derive a weighted mean of: $\Gamma_{\rm eff}=$\,1.98\,$\pm$\,0.07, in reasonable agreement with the expected value.

\addtocounter{figure}{1}
%
%
 \begin{figure*}[tbh]
 \centerline{\psfig{file=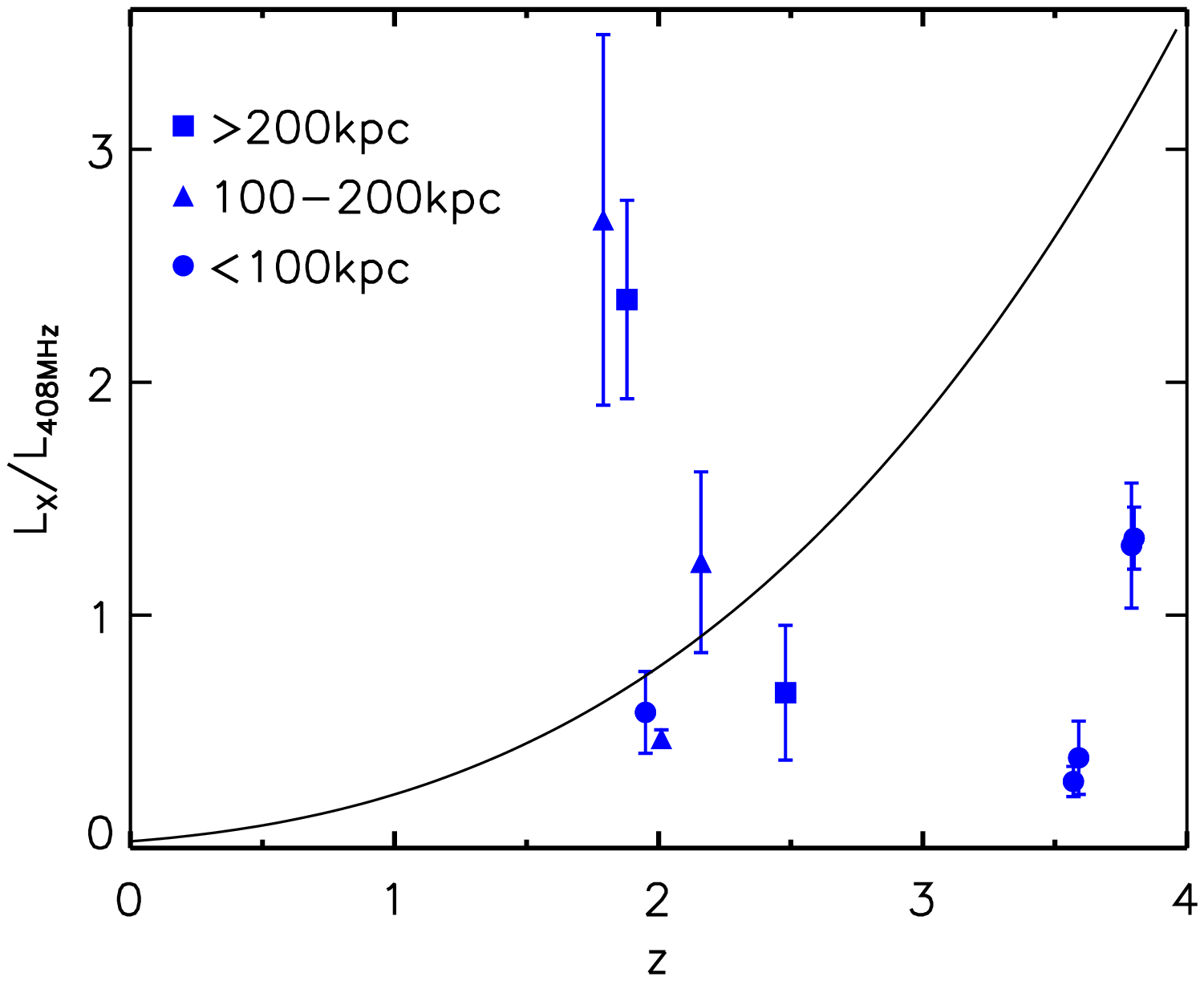,width=3.2in,angle=0} \psfig{file=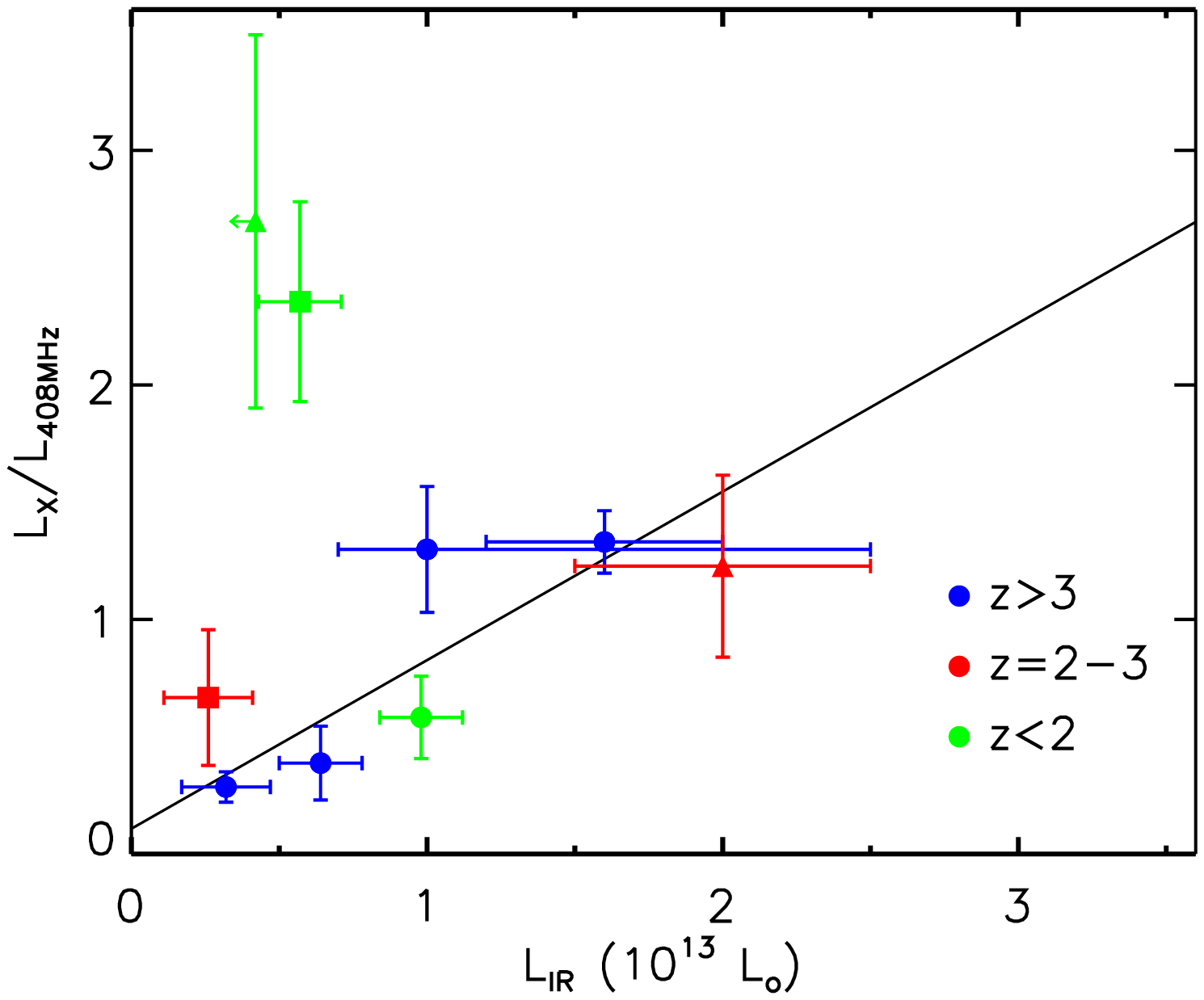,width=3.2in,angle=0}}
 \vspace*{-0.5cm}
 \centerline{\psfig{file=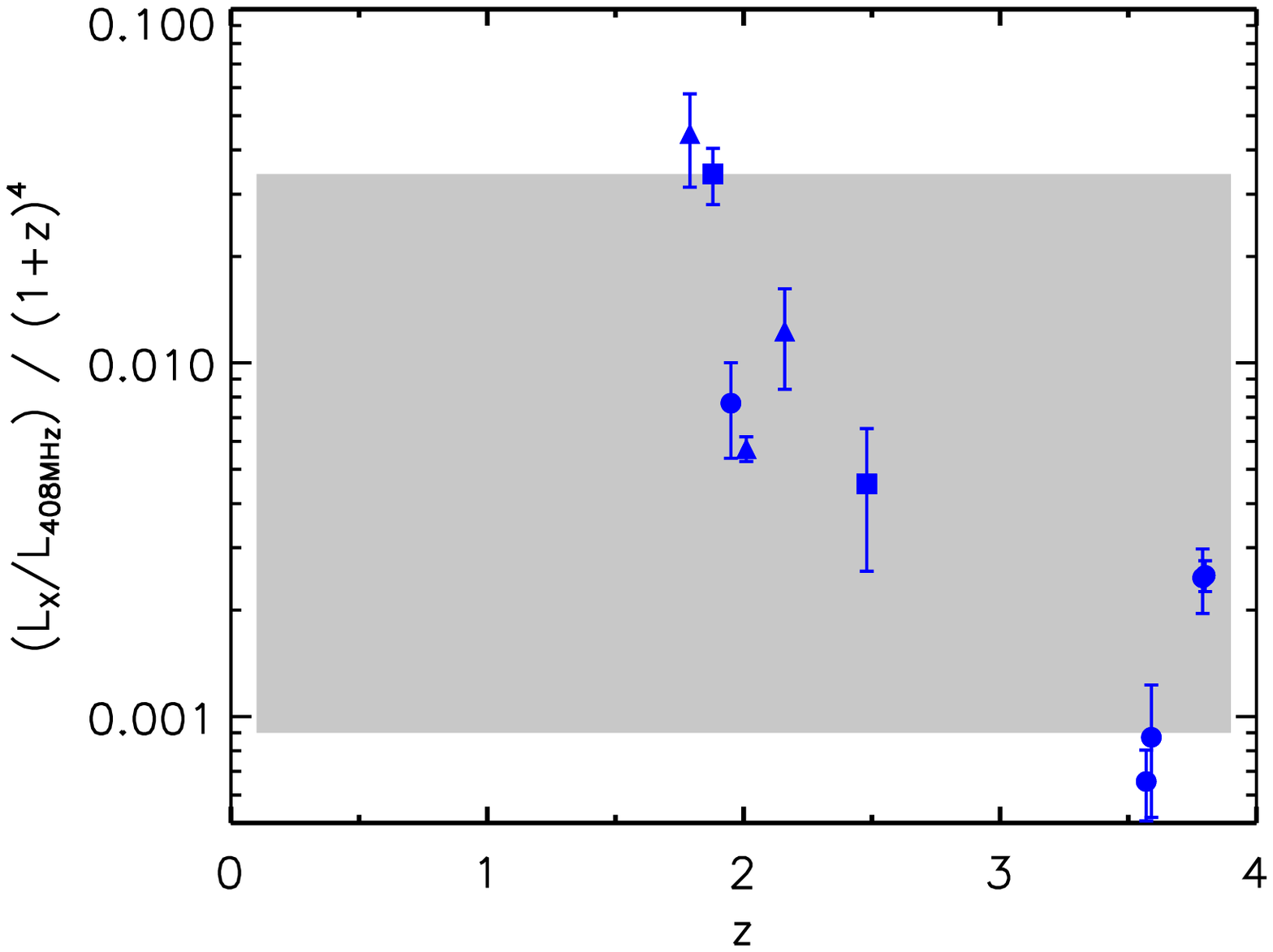,width=3.2in,angle=0} \psfig{file=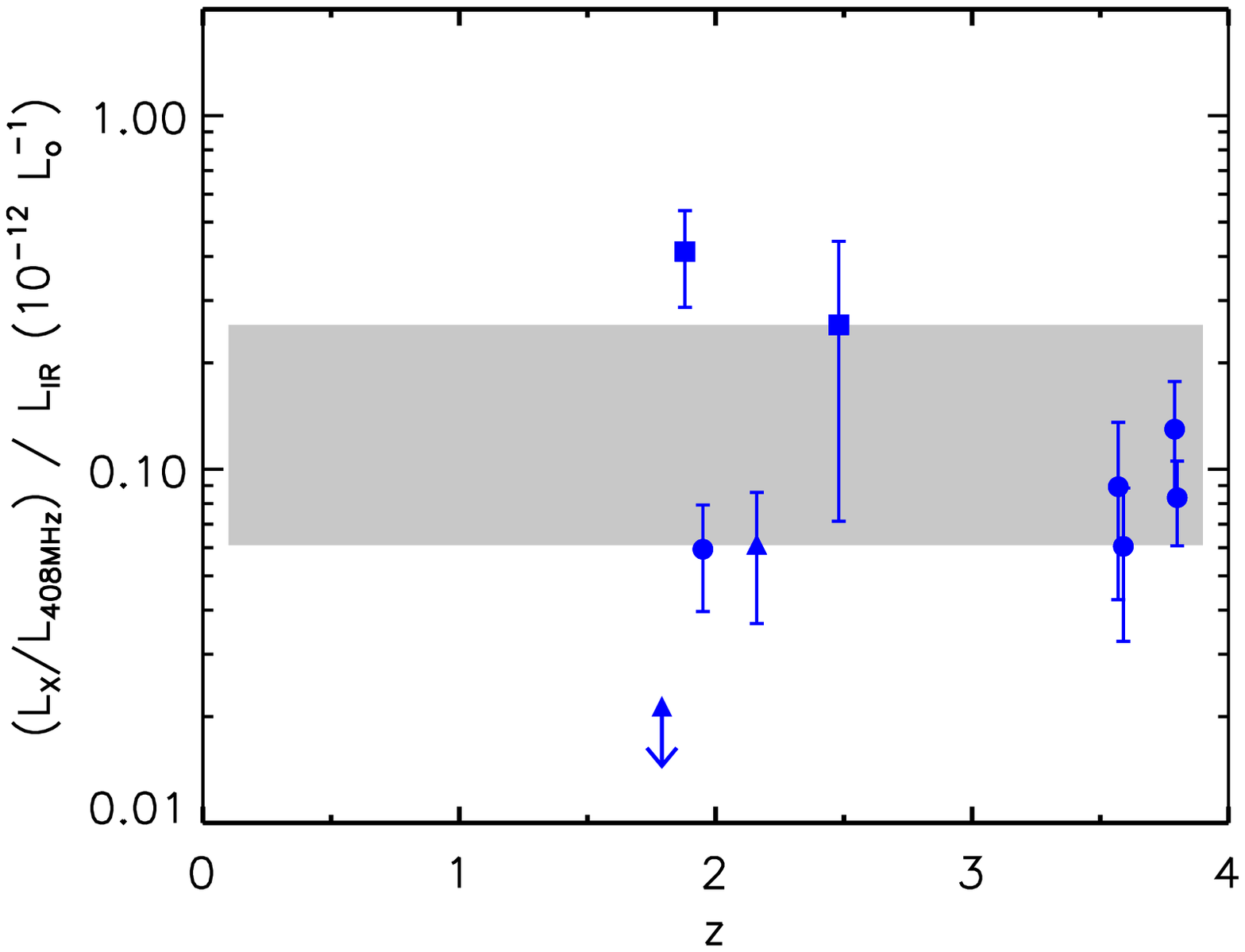,width=3.2in,angle=0}}

 \vspace*{-0.3cm}
\caption{\small {\it (upper-left)} The variation in X-ray to radio luminosity versus redshift for the high-redshift radio sources with  IC-powered X-ray emission from Table~2.  We code the points on the physical extent of the radio lobes in Table~2. We plot over this a curve illustrating the behaviour expected if the IC emission is being driven by the CMB photon field, whose density rises as $(1+z)^4$, and if all the sources have comparable  ages and jet powers. There is little indication that the data follow this trend (and no obvious systematic variation with radio source size), either because the CMB is not the relevant source of photons for the IC scattering, or because other relevant physical parameters vary significantly between sources (see Mocz et al.\ 2011a,b).  
{\it (upper-right)} The variation of X-ray to radio luminosity ratio versus total infrared luminosity for our high-redshift radio sources  (now coded by radio lobe size and redshift).  We see weak evidence for a correlation between the IC emission and the total infrared luminosity of the radio sources: a fit to all the data yields a near-linear gradient of 0.72\,$\pm$\,0.10, although this is a poor fit due to the two $z<$\,2 outliers (removing these yields a gradient of 0.75\,$\pm$\,0.10 and a reduced $\chi^2\sim $\,1).  
{\it (lower-left)} The variation in $L_{\rm X}/L_{\rm 408\,MHz}$ versus redshift for the radio sources, after normalising the observed frame X-ray-to-radio ratio by the expected redshift evolution due to the CMB photon density: $(1+z)^4$. Note the more than two orders of magnitude scatter between sources, the lack of any correlation with  source size and the the spurious trend with redshift.  The gray region indicates the 2-$\sigma$ confidence limits on the median of the distribution derived from bootstrap resampling.  
{\it (lower-right)} The same X-ray-to-radio luminosity ratio, now normalised by the total infrared luminosities of the radio sources, assuming that these provide the seed photons for the IC emission.  The dispersion is reduced by almost $\sim $\,4$\times$ compared to the left-hand panel, as illustrated by the gray region.  We code the galaxies by the extent of their radio lobes and we note that the two largest radio sources also show the  largest far-infrared normalised X-ray-to-radio ratios.  If we remove these systems on the assumption that they have a significant CMB contribution to their $\sim$\,0.5--1-Mpc IC emission, then the remaining radio sources show just a $\sim$\,20\% scatter, which is consistent with their measurement errors.  We plot the observation of 3C\,294 as a 3-$\sigma$ limit and do not show 3C\,9 for which there is no constraint on the infrared luminosity.  }
\end{figure*}

The spectral properties of the X-ray emission are thus roughly consistent with the expectation for IC scattering of either far-infrared or CMB photons off electrons with $\gamma\sim$\,1000s.  There are three broad tests which we can apply to try to distinguish between far-infrared and CMB sources: (1) using the spatial or morphological properties of the X-ray emission, (2) by identifying a correlation of the IC emission with either redshift, or (3) with total infrared luminosity of the radio sources.    Note that these last two may be hard to decouple from other evolutionary effects, such as source age and lobe expansion.  

We start by investigating the radial profile of the X-ray emission around the HzRGs to see what information it is possible to discern on the source of the photon field responsible for the IC emission.  As noted above, the correspondence between the X-ray emission and the radio lobe hotspots in 4C\,19.71 maybe hinting at the dominance of CMB-powered IC emission in this far-infrared-faint HzRG.  However, the halos around the two $z\sim$\,3.6 HzRGs are unfortunately too faint to provide reliable individual surface brightness profiles.  Instead, we can take advantage of the homogeneity of our sample by stacking the four $z>$\,3 HzRGs: 4C\,03.24, 4C\,19.71, 4C\,60.07 and 4C\,41.17 (although we caution that 4C\,41.17 dominates the combined flux owing to its brightness).  We rotate the X-ray images to align the major axes of the radio lobes in the galaxies, centroid and coadd them.  We then extract the radial surface brightness profile, having  masked the core emission by interpolation, and show this in Fig.~3.  We fit a power law to the surface brightness profile and derive a power-law index of $n \sim -$1.2 (Fig.~3).  We also plot on the profile expected for a toy model with a central source illuminating a uniform cylindrical radio lobe, with an axial ratio of 10:3 (similar to the X-ray morphology of 4C\,41.17).  

As can be seen in Fig.~3, the average surface brightness profile of the IC halos is consistent either with a highly-collimated electron population illuminated by a photon field which is more extended than a central point source, or a relatively ``fat'' electron reservoir, with an axial ratio of $\sim$\,10:3, but in this case illuminated by a central source.  We conclude that the spatial profile of the X-ray emission of the IC halos is currently too poorly constrained to rule out either of the potential sources of photons, and significantly deeper observations would be required to reliably apply this test.   Nevertheless, as we show, separating the sources  based on the extent of their radio lobes may provide an indication of the relative contributions of the two photon sources, as it is unlikely that far-infrared photons could dominate the IC scattering in the largest, $\gs$\,500\,kpc, lobes (Laskar et al.\ 2010).

As our next test of the origin of the IC emission we investigate the variation of the IC emission with redshift.  The IC emission can be expressed as the ratio of extended X-ray to radio luminosities, which in a naive model of a non-evolving radio galaxy should only depend on the far-infrared/CMB photon density and the $B$-field: 

$$L_{\rm X}/L_{\rm 408MHz} = 8 \pi \rho / B^2 \propto (1+z)^4 $$  

We plot the variation of this quantity with redshift in Fig.~4.  As can be seen, there is little evidence for the strong $(1+z)^4$ rise expected if $\rho$ is dominated by the CMB. To quantify this, we assume the CMB is the dominant source of photons and divide through by $(1+z)^4$ and plot the behaviour of this quantity versus redshift in the lower-left panel. Instead of yielding a constant value, this now shows over a two  orders of magnitude decline across $z$\,$\sim$\,2--4, and we highlight the behaviour of the four $z$\,$=$\,3.6--3.8 HzRGs, which show a variation of a factor of $\sim$\,4 over a negligible range in redshift.   We also identify the  different physical scales of the sources IC X-ray emission in these two panels, based on the lobe sizes listed in Table~2.    We see no obvious trend in X-ray to radio luminosity ratio when separating the sample on IC X-ray halo extent, as expected from the analysis of Mocz et al.\ (2011a,b).   We conclude that if the CMB is the primary source of photons driving the IC emission in these radio sources, then there must be other physical factors which dominate its evolution, other than the rapid rise in the CMB density with redshift.  The latter is perhaps not unexpected, as discussed by Mocz et al.\ (2011a), given that the  X-ray to radio luminosity ratio depends on a number of factors including the bulk kinetic power transported along the jets to fuel the lobes (which governs the magnetic field in the lobes), the low-energy turnover (of the energy distribution of the electrons available for up-scattering photons), as well as the epoch at which the radio galaxy is observed, due to the evolution and expansion of radio synchrotron lobes causing their magnetic field to fall and the energy distribution to rapidly shift to lower energies.

As our final test we turn to the other potential contributor to the scattering photon field responsible for the IC: far-infrared emission from dust-obscured starbursts in these galaxies.  To test this hypothesis we plot $L_{\rm X}/L_{\rm 408MHz}$ versus $L_{\rm IR}$ in Fig.~4.  This shows a modest correlation, which is more obvious in the $z$\,$>$\,2 radio sources.  A fit to all the data gives a gradient of 0.72\,$\pm$\,0.10, broadly consistent with a linear correlation, although the fit is poor due to the two $z$\,$<$\,2 outliers. Removing these two outliers gives a gradient of 0.75\,$\pm$\,0.10 and a reduced $\chi^2\sim $\,1.  To compare the dispersion in this model to that for the CMB, we plot $L_{\rm X}/(L_{\rm 408MHz}\times L_{\rm IR})$ versus redshift in the lower-right panel of Fig.~4.  This figure shows an order of magnitude lower dispersion than the equivalent CMB plot and no obvious trend with redshift: in particular the four $z=$\,3.6--3.8 HzRGs are all consistent with each other.  In addition, the two HzRGs with the largest lobe extents are also outliers on this plot.  These two HzRGs have lobe lengths of $\sim$\,0.5--1\,Mpc, sufficiently large that it is unfeasible that the far-infrared emission will dominate over the CMB at such large distances. Removing these galaxies from Fig.~4, we find that the scatter in the remainder of the sample is consistent with the measurement errors at just $\sim $\,20\%.  This suggests that at least in the majority of more compact HzRGs, their luminous, dusty starbursts may be a significant source of the photons driving their IC emission.  

As well as these trends, we can also ask if the far-infrared emission is sufficient to power the IC X-ray emission.  Strong evolution is seen in the total infrared luminosities of HzRGs (Archibald et al.\ 2001). There is also growing evidence for extended far-infrared emission around some HzRGs (e.g.\ Stevens et al.\ 2003; Ivison et al.\ 2008, 2010, 2012).  We can now estimate at what radius the far-infrared emission from the starburst will exceed the CMB luminosity density, which at $z=$\,3.6--3.8 is $\rho_{\rm CMB}\sim $\,2\,$\times $\,10$^{-10}$\,erg\,cm$^{-3}$.  The far-infrared energy density from a uniform-density, spherical emission distribution around the HzRGs is $\rho_{\rm IR}\sim $\,9\,$L_{\rm IR}/2\pi\,c\,R^2$ (Scharf et al.\ 2003).  For the observed total infrared luminosities, we find that $\rho_{\rm IR}=\rho_{\rm CMB}$ occurs at radii $\sim$\,20--25\,kpc for 4C\,03.24 and 4C\,19.71 (c.f.\ 30--40\,kpc for 4C\,60.07 and 4C\,41.17). These scales are well-matched to the sizes of the radio and X-ray emission around these $z>3$ galaxies (radii of 20--40\,kpc, Fig.~1 \& 2).

Hence, the far-infrared emission in these two HzRGs appears to be sufficient to power the IC X-ray emission, as concluded previously by Scharf et al.\ (2003) and Smail et al.\ (2009) for 4C\,41.17 and 4C\,60.07.  Moreover, the spatial scale on which the far-infrared emission from the HzRGs could dominate over the CMB in these compact radio sources is comparable to that of the X-ray and ionised halos. Indeed, the correlation of the IC X-ray emission with the Ly$\alpha$ or [O{\sc iii}] emission-line morphology suggests that this process is indeed having an impact on the gas reservoirs around these massive galaxies.  The majority of the radio sources  (8/10) in the $z\gs$\,2 sample we have analysed have X-ray IC emission on scales (and radio lobe extents) of $\ls $\,200\,kpc.  This is also the region around these galaxies within which significant quantities of neutral gas are likely to be found (van Ojik et al.\ 1997) and so the resulting IC-powered X-ray emission provides a natural, extended source of direct ionisation needed to create the large, kinematically quiescent Ly$\alpha$ emission-line halos frequently associated with such systems (Villar-Mart\'in et al.\ 2003).  This process may act in addition to the photoionisation by the quasar nucleus and shock heating, resulting from the passage of the radio lobes, which appear to be the dominant mechanisms heating the typically less-extended, but higher surface brightness, ionised structures seen in radio galaxies (e.g.\ Villar-Mart\'in et al.\ 1999; Best et al.\ 2000; Tadhunter et al.\ 2000)

If the far-infrared emission is the dominant driver of the evolution of the IC X-ray emission in young, compact radio sources, we can draw two conclusions: As the total infrared luminosities of HzRGs evolves more slowly than the CMB (Archibald et al.\ 2001): $(1+z)^3$ versus $(1+z)^4$, then at some redshift the CMB will eventually exceed the total infrared luminosities of the HzRGs even on small spatial scales.  Pushing observations of IC emission around HzRGs to even higher redshifts, $z\gs$\,4, in both far-infrared-bright and -faint HzRGs is thus an important test of this hypothesis. More critically, the fact that the IC emission is driven by the far-infrared emission from the HzRGs means that this feedback mechanism is dictated entirely by {\it local} processes, rather than the cosmological evolution of the CMB.  

The requirement to have both intense radio jet emission from an AGN, synchronised with similarly intense far-infrared emission, might be thought to limit the prevalence of this mechanism.  However, as  Blundell \& Rawlings (1999) pointed out that there is a strong and inevitable bias towards only detecting the youngest radio galaxies at higher redshifts.  This arises because the luminosities of  radio lobes decline rapidly with age, and thus these sources are most easily detectable in their first flush of youth.   These high-redshift radio sources are also those which are expected to have the most intense starburst activity (e.g.\ Archibald et al.\ 2001; Dey et al.\ 1997). Thus we conjecture that it is in these compact ($\ls $\,200\,kpc) radio galaxies where the far-infrared-derived IC will dominate due to the presence of intense starburst activity, while in the largest ($\gg$\,200\,kpc) HzRGs at $z \ls $\,2, where aged radio emission lies significantly beyond the sphere of influence of any starburst,  the CMB will provide the dominant  photon field.    Moreover, if radio sources are prone to repeated episodes of jet formation (e.g.\ Reynolds \& Begelman 1997; Nipoti et al.\ 2005; Blundell \& Fabian 2011), then it is likely that the far-infrared-driven mode of IC halo formation will be enhanced.  This is because the electron populations from previous phases of activity (seen as aged or relic lobes, or indeed invisible in current radio studies, e.g.\ Fabian et al.\ 2009) will contribute significantly to the population of $\gamma \sim $\,100 particles. Support for this intrepretation comes from the off-axis X-ray emission of 3C\,294 (Erlund et al.\ 2006), as well as similar misalignments seen between the radio and X-ray IC emission and spectral aging in other sources (e.g.\ Jamrozy et al.\ 2007).  This leaves open the possibility that far-infrared-enhanced IC emission from luminous starbursts, which are believed to be relatively long-lived compared to luminous radio jet events, combined with the large (and less variable) populations of low-$\gamma$ electrons produced by repeated radio activity, may mean that IC X-ray halos are more prevalent than currently predicted (e.g.\ Mocz et al.\ 2011a,b). Indeed, the radio-lobe-less IC X-ray halo around HDF\,130 (Fabian et al.\ 2009) may be an example of such a system as {\it Herschel} SPIRE observations detect far-infared emission with a luminosity of $\sim$\,10$^{13}$\,L$_\odot$ from this galaxy.

\section{Conclusions}

The main conclusions of this work are:

1. We detect faint, extended X-ray emission around 4C\,03.24 and 4C\,19.71: two powerful radio galaxies at $z\sim$\,3.6.  The extended X-ray emission halos have luminosities of $L_{\rm X}\sim$\,3\,$\times$\,10$^{44}$\,erg\,s$^{-1}$ and linear extents of $\sim $\,60\,kpc.  These detections double the number of $z$\,$>$\,3 radio galaxies with extended luminous X-ray halos.

2. The spatial distributions of the X-ray emission are broadly similar to the radio emission from these galaxies and so they are likely to be formed by IC scattering of either CMB photons or locally-produced, far-infrared photons by the radio lobes' relativistic electrons.

3. If the photon field powering the X-ray emission in these sources is the CMB, then as a result of the $(1+z)^4$ dependence of the CMB photon density, and assuming the ages and jets powers of the radio galaxies are identical, we would expect the two newly-discovered X-ray halos to be $\sim $\,20\% less luminous than the X-ray halos previously detected around two comparably radio-luminous galaxies at $z$\,$\sim$\,3.8.  However, the two newly-discovered X-ray halos are in fact both $\sim$\,4\,$\times$ less luminous than those previously found.  We interpret this as evidence that either the CMB is not providing the dominant photon population responsible for the IC scattering in these galaxies, or that their magnetic fields or the electron populations within the radio lobes are rapidly evolving, in a manner which negates the expected strong redshift evolution.

4.  We suggest a more plausible explanation for the faint IC X-ray halos around the two $z\sim $\,3.6 HzRGs is that the dominant seeds for the IC emission in these compact radio sources are not millimeter photons from the CMB, but instead are far-infrared photons which are produced in dust-obscured starbursts within these galaxies.  The total infrared luminosities of the two $z$\,$\sim$\,3.6 radio galaxies are $\sim$\,4\,$\times$ lower than the two $z$\,$\sim$\,3.8 examples, matching the observed difference between their IC X-ray luminosities.  We show further evidence for a correlation between the total infrared luminosity and the IC X-ray emission around a larger sample of radio sources from the literature, especially when restricted to those  with radio lobe lengths of $\ls $\,200\,kpc.  However, we caution that this literature sample is not complete and that may be examples of far-infrared bright, high-redshift radio sources where IC emission is not detected in sensitive X-ray observations, but which as a result have not been published.

We conclude that, depending upon the redshift, total infrared luminosity and physical extent of the radio lobes, it is possible that a mix of both locally produced far-infrared and CMB photons is responsible for the IC emission from high-redshift radio sources.  These alternative sources of seed photons may help explain the significant scatter seen in the strength of IC X-ray emission from these systems.  The cosmological significance of including  far-infrared photons as a potentially dominant photon source for IC emission around compact HzRGs, rather than the CMB, arises from the fact that this means this is an entirely ``local'' process, employing locally-produced synchrotron electrons and locally-produced far-infrared photons from a dusty starburst.  The  intense X-ray emission which results provides a powerful feedback mechanism, sufficient to ionise the gaseous halo of the galaxies on $\sim $\,100-kpc scales and forming extended emission line halos around the galaxies (as seen around most of the examples discussed here). Thus the starburst and AGN activity are feeding off each other to produce a more effective and potentially more wide-spread  {\it combined} feedback process than either could individually.  Moreover, if episodic radio activity is common and roughly synchronised with intense starburst activity in massive, high-redshift galaxies, then this far-infrared-driven IC-feedback mechanism could have a significant role in affecting the star-formation histories of the most massive galaxies at the present-day.   Although we note that not all these conditions may be met, as not all quasar activity is  associated with powerful radio jets, nor with a  coeval starburst (Hickox et al.\ 2012).  Finally, by breaking the tie to the CMB photon density, such IC-powered halos may be more prevalent than previously predicted (e.g.\ Mocz et al.\ 2011a) and also need not be restricted to the highest redshifts, although the $(1+z)^3$ evolution of the characteristic total infrared luminosity of HzRGs (Archibald et al.\ 2001) would suggest the prevalence of such X-ray halos may increase with redshift as the frequency and strength of their starburst activity increases.

\acknowledgments 
We thank for Nick Seymour for generously checking his data prior to publication and Rob Ivison, Richard Bower, Caleb Scharf, Wil van Breugel, Mark Dickinson, Arjun Dey, Hy Spinrad and Dan Stern for help and useful conversations.  I.R.S.\ acknowledges a Leverhulme Trust  Fellowship and I.R.S., K.M.B.\ and D.M.A.\ acknowledge support from the STFC.  B.D.L.\ acknowledges support from an Einstein Fellowship and {\it CXC} grant G01-12173X.  This work has used data from the NASA Extragalactic Database (NED), and from the NRAO VLA, {\it Hubble Space Telescope} and {\it Spitzer Space Telescope} data archives.

\setlength{\tabcolsep}{3pt} 

%
%
\begin{center}{\small
\begin{table}[ht]
\small
\centerline{\sc Table 1}
\centerline{\sc Properties of HzRGs}
\smallskip
\centerline{\begin{tabular}{cccccccl}
\hline
ID & $z$ & $L_{\rm IR}$ (8--1000\,$\mu$m) & $L_{\rm 1.7\,GHz}$ & $L_{\rm X}$ (0.5--8\,KeV)  & $\Gamma_{\rm eff}$ &  $L_{\rm Ly\alpha}$  & Comment\\
   &      & (10$^{12}$\,L$_\odot$) & (10$^{44}$\,erg\,s$^{-1}$) & (10$^{44}$\,erg\,s$^{-1}$) &  & (10$^{44}$\,erg\,s$^{-1}$)   & \\
\hline\hline
4C\,03.24 & 3.57 & $6.4\pm 1.4$ & $1.9\pm 0.1$ & $3.4^{+0.9}_{-0.8}$ & 1.8$^{+0.7}_{-0.5}$ & 3.0 & This paper  \\  
4C\,19.71 & 3.59 & $3.2\pm 1.5$ & $1.5\pm 0.1$ & $3.8^{+1.5}_{-1.2}$ & 1.1$^{+0.9}_{-0.6}$ & 2.1 & This paper  \\
\noalign{\smallskip}  
4C\,60.07 & 3.79 & $10^{+15}_{-3}$ & $1.4\pm 0.1$ & $12\pm 2$ & $0.8^{+0.6}_{-0.7}$ & 12 & Smail et al.\ (2009)  \\  
4C\,41.17 & 3.80 & $16\pm 4$ & $1.4\pm 0.1$ & $12\pm 3$ & $1.6^{+0.3}_{-0.3}$ & 12 & Scharf et al.\ (2003)  \\  
\hline
\end{tabular}}
\smallskip
\end{table}}
\end{center}
\vspace*{-0.5cm}

%
%
\begin{center}{\small
\begin{table}[ht]
\small
\centerline{\sc Table 2}
\centerline{\sc Properties of HzRG with published IC detections}
\smallskip
\centerline{\begin{tabular}{ccccccccccl}
\hline
ID & $z$ & $f_{\rm X}$ (0.5--8\,KeV) & $\Gamma_{\rm eff}$  &$f_{\rm 408\,MHz}$ & $f_{\rm 850\mu m}$ & $L_{\rm X}$\,(0.5--8\,KeV) & $L_{\rm 408\,MHz}$ & $L_{\rm IR}$  & D$_{\rm Lobe}$ &  Ref.\\
   &      & (10$^{-15}$\,erg\,cm$^{-2}$\,s$^{-1}$) & & (Jy) & (mJy) & (10$^{11}$\,L$_\odot$) & (10$^{11}$\,L$_\odot$) &  (10$^{12}$\,L$_\odot$) &   (kpc) & \\
\hline\hline
3C\,294   & 1.79 & $70 \pm 20$ & $2.3 \pm 0.1$ & $5.2 \pm 0.4$ & $0.2\pm 0.8$ &$3.8 \pm 1.1$ & $1.42 \pm 0.11$ & $<$\,4.2 & 130 & E06, A01\\
6C\,0905  & 1.88 & $11 \pm 2$ & $1.6 \pm 0.2$ & $0.94 \pm 0.03$ & $3.6\pm 0.9$  &$0.73 \pm 0.13$ & $0.31 \pm 0.01$ & $5.7\pm 1.4$ & 950 & E08, A01\\
3C\,191   & 1.95 & --- & $1.6 \pm 0.4$ & $7.3 \pm 0.3$ & $6.4\pm 1.1$  &$6.0 \pm 1.8$ & $10.3 \pm 0.3$ & $9.8\pm 1.4$ & 40 &E06\\
3C\,9     & 2.01 & $19 \pm 0.5$ & $1.6 \pm 0.6$ & $7.8 \pm 0.3$ & ---  & $1.4 \pm 0.1$&$2.98 \pm 0.11$ & --- & 120 & F03 \\
MRC\,1138 & 2.16 & $26 \pm 8$ & $1.8 \pm 0.2$ & $4.12 \pm 0.11$ &  $13\pm 3$  & $5.4 \pm 1.7$& $4.40 \pm 0.12$ & $20\pm 5$& 130 &O05, R04\\
4C\,23.56 & 2.48 & $5.3 \pm 2.0$ & $3.2 \pm 1.0$ & $1.60 \pm 0.10$ & $1.7\pm 1.0$  & $0.7 \pm 0.3$&$1.05 \pm 0.07$ & $2.6\pm 1.5$& 490 & J07, A01\\
4C\,03.24 & 3.57 & $2.9 \pm 0.7$ & $1.8 \pm 0.7$ & $1.81\pm 0.08$ & $2.3\pm 1.1$  & $0.9 \pm 0.2$&$3.15 \pm 0.14$ & $3.2\pm 1.5$& 60 &S12, A01\\
4C\,19.71 & 3.59 & $3.2 \pm 1.3$ & $1.1 \pm 0.9$ & $1.46\pm 0.10$ & $4.6\pm 1.0$  & $1.0 \pm 0.4$& $2.58 \pm 0.18$& $6.4\pm 1.4$& 60& S12, A01\\
4C\,60.07 & 3.79 & $8.2 \pm 1.6$ & $0.8 \pm 0.4$ & $1.13\pm 0.06$ & $17.1\pm 1.3$  & $3.0 \pm 0.6$& $2.31 \pm 0.12$& $10^{+15}_{-3}$& 65 & S09, A01\\
4C\,41.17 & 3.80 & $8.5 \pm 0.9$ & $1.6 \pm 0.3$ & $1.14 \pm 0.03$ & $12.1\pm 0.9$  & $3.1 \pm 0.3$& $2.33 \pm 0.06$& $16\pm 4$& 95& S03, A01\\
\hline
\end{tabular}}
\smallskip
{\small  Notes: Radio fluxes and redshifts are from literature sources compiled by NED. X-ray fluxes and spectral indices are
from:  
E06, Erlund et al.\ (2006);
E08, Erlund et al.\ (2008);
F03, Fabian et al.\ (2003);
J07, Johnson et al.\ (2007);
O05, Overzier et al.\ (2005);
 S12, this work;
S09, Smail et al.\ (2009); S03, Scharf et al.\ (2003).  Sub-millimeter photometry comes from: A01, Archibald et al.\ (2001); R04, Reuland et al.\ (2004).  Fluxes are in observed bands, luminosities are restframe.}
\end{table}}
\end{center}
\end{document}